# Synthesis, Characterization, and Finite Size Effects on Electrical Transport of Nanoribbons of the Charge-Density Wave Conductor $NbSe_3$

**Adam A Stabile[1,#], Luisa Whittaker[2,#], Tai Lung Wu[1], Peter Marley, Sarbajit Banerjee[2,*] and G. Sambandamurthy[1,*]**

[1]Department of Physics, University at Buffalo, Buffalo, New York 14260-1500
[2]Department of Chemistry, University at Buffalo, Buffalo, New York 14260-3000

* Email: sb244@buffalo.edu and sg82@buffalo.edu
[#] These authors contributed equally to this work.

**Abstract**

$NbSe_3$ exhibits remarkable anisotropy in most of its physical properties and has been a model system for studies of quasi-one-dimensional charge-density-wave (CDW) phenomena. Herein, we report the synthesis, characterization, and electrical transport of single-crystalline $NbSe_3$ nanoribbons by a facile one-step vapour transport process involving the transport of selenium powder onto a niobium foil substrate. Our investigations aid the understanding of the CDW nature of $NbSe_3$ and the growth process of the material. They also indicate that $NbSe_3$ nanoribbons have enhanced CDW properties compared to those of the bulk phase due to size confinement effects, thus expanding the search for new mesoscopic phenomena at the nanoscale level. Single nanoribbon measurements on the electrical resistance as a function of temperature show charge-density wave transitions at 59 K and 141 K. We also demonstrate significant enhancement in the depinning effect and sliding regimes mainly attributed to finite size effects.

## 1. Introduction

Charge-density-wave (CDW), a phenomenon characteristic of quasi-one dimensional correlated electronic systems, wherein a Fermi sea of charge carriers modulates with respect to a distorted underlying lattice, has garnered much interest as a rather eclectic example of the coupling of electronic and geometric structure in the solid-state [1, 2]. Much interest in CDW properties derives from the close analogies to superconducting materials, namely pronounced electron-phonon interactions and the formation of an energy gap [3]. In fact, signatures of CDW properties have been seen to exist in many known superconducting materials [4, 5]. Recently, it has been shown that certain materials known for exhibiting charge-density-wave behaviour can also show superconductivity simply by tuning the concentration of specific dopants [6]. CDW materials have further been used to create MOSFET heterostructures, thus making them suitable candidates for electronic applications in the semiconducting industry [7].

Since CDWs are observed in materials with a quasi-one dimensional or chain-like atomic structure, reducing their dimensions portends the likelihood of the manifestation of unique phenomena with considerable scope for the manipulation of electronic instabilities. $NbSe_3$ is a textbook example of CDW material, and its properties in bulk form have been known for decades [8]. However, finite size effects on $NbSe_3$ have not been as comprehensively explored [9, 10]. Thus far, top-down approaches have been the mainstay for investigating nanoscale CDW materials with the adoption of a two-step approach wherein a macroscopic crystal is created using the reliable and widely known method of chemical vapour transport (CVT) [11], and then nanofabrication techniques such focused ion beam (FIB) etching or ultrasonic cleaving [12] are employed to scale the materials to nanoscopic dimensions. A critical gap in our knowledge of these materials is the behaviour of CDW nanostructures prepared by bottom-up approaches from elemental or molecular precursors, which are expected to exhibit unique finite size effects that result not just from geometric confinement but also the lattice reconstruction and unique surface termination adopted to minimize energy at nanoscale dimensions [13]. Furthermore, top down approaches often degrade the crystal quality; specifically, gallium ions from the FIB may add undesired impurities to the sample or destroy the crystallinity completely as seen in other materials [14].

Novel techniques for the fabrication of $NbSe_3$ nanoribbons by bottom-up approaches have appeared in literature [15]. We report a new, facile approach to CVT, which can yield single-crystalline nanostructured samples in a single step. We begin with a description of the synthetic techniques followed by a discussion of the dimensions, morphology, and composition of our nanoribbons. Lastly, we describe its electrical transport properties, which exhibit characteristic behaviour of nanoscale $NbSe_3$.

## 2. Experimental Section

### *2.1 Synthesis of $NbSe_3$ Nanoribbons*

$NbSe_3$ nanoribbons were synthesized from elemental selenium (Se) and a metallic niobium (Nb) foil by CVT in a one-step reaction. We start with a fused-silica tube, which has a width constriction 50 mm along its length. As shown in Figure 1(a) a piece of Nb foil and elemental powders of Se are placed at opposite ends of this constriction and the tube is evacuated to ~$10^{-3}$ Torr on a vacuum line prior to sealing. The 50 mm separation and induced constriction in the tube prevents the precursors from coming into direct physical contact as the tube was sealed or during the reaction. The fused-silica tube was heated in a horizontal tube furnace to 700 °C for 2 h enabling sublimation and transport of elemental Se to the Nb foil as per an established diffusion gradient. A dense formation of nanoribbons on the Nb substrate was observed and further characterized after removal from the substrate.

*2.2 Characterization*

Phase identification and evaluation of the crystallinity of the as-deposited $NbSe_3$ nanoribbons were performed by using X-ray diffraction (XRD) on a Rigaku Ultima IV diffractometer at a scanning rate of 3° per minute in the 2θ range between 5 and 65° using graphite-monochromated Cu $K_\alpha$ radiation ($\lambda = 0.15418$ nm) at 40 kV accelerating voltage and 40 mA current.

The dimensions, morphology, and elemental composition of the as-deposited $NbSe_3$ nanoribbons were examined by means of scanning electron microscopy (SEM) and energy-dispersive X-ray (EDX) spectroscopy using a Hitachi SU-70 SEM operating at an accelerating voltage of 20 keV. A JEOL-2010 instrument operating at 200 keV was used to acquire high-resolution transmission electron microscopy (HRTEM) images and selected-area electron diffraction (SAED) patterns. To prepare the samples for HRTEM/SAED analysis, the nanostructures were dispersed in 2-propanol and then deposited onto 300 mesh carbon-coated Cu grids.

*2.3 Electrical Transport Measurements*

As-prepared $NbSe_3$ nanoribbons deposited on the Nb foil were initially immersed in 2-propanol allowing small bundles of nanoribbons to be detached from the substrate. In order to separate them into single nanoribbons, the solution was ultrasonicated for a few seconds, and further diluted with 2-propanol. The supernatant was extracted with a syringe, dropped onto a Si/$SiO_2$ substrate, and then blown dry with $N_2$ gas. Contacts were patterned on the thinnest wires (roughly identified by optical contrast against the 300 nm $SiO_2$ surface) by optical or by electron-beam lithography, and metalized with 5 nm of Cr and with 75 nm of Au in an e-beam evaporator. Transport measurements were carried out in Variable Temperature Insert using standard, low-frequency lock-in techniques. These techniques include four-terminal or two-terminal contact configurations with an AC excitation current of 100 nA or lower.

## 3. Results and Discussion

$NbSe_3$ crystallizes in the monoclinic $P2_1/m$ space group (figure 1(b)). This structure consists of columns of face-sharing triangular prisms of Se atoms, aligned parallel to the monoclinic *b* axis, within which a single Nb atom is trigonal prismatically coordinated by six Se atoms. The $NbSe_3$ unit cell consists of three types of symmetry-related independent chain types forming corrugated layers held together by van der Waals forces [16]. As labelled in figure 1(c), the three chains are differentiated by the outer Se-Se bond length in each trigonal prism with bond lengths of 2.484, 2.909, and 2.378 Å, respectively [16]. The bond strengths play a noteworthy role in determining the local charge density, directional anisotropy in electrical conduction, and the charge-density-wave properties, and may perhaps be amenable to subtle modification as a result of scaling to finite size [10, 16, 17].

Figure 2 depicts a typical SEM micrograph of the as-deposited $NbSe_3$ nanostructures which tend to crystallize as anisotropic ribbons (and not cylindrical wires). It is worth mentioning that the $NbSe_3$ crystal structure has two easy cleavage planes: one parallel to the crystallographic *bc* plane and the second parallel to the *ab* plane, both of which ease the reduction of ribbon-like structures into thin nanofibers. In all cases, we note the formation of highly faceted nanoribbon structures with approximately rectangular cross-sections, as evidenced from cross-sectional high-resolution SEM images shown in figure 2(a). The nanoribbons show a distribution of widths centered at 198 ± 53 nm (based on measurements of 50 nanoribbons, Fig. 3e) but even the thinner nanowires have still smaller vertical heights and thus rectangular cross-sections as depicted in Fig. 3. The EDX spectrum presented in figure 2(b) corroborates the presence of Nb and Se atoms in the nanoribbons with an elemental composition of $Nb_{25.68}Se_{74.32}$ deduced for the nanowires, which corresponds well to the ideal $NbSe_3$ geometry within the limits of experimental error (approx. 1 at.%). No discernible concentration gradients were observed while mapping the elemental composition along the length of the nanoribbons. Figure 2(c) illustrates XRD patterns acquired for the as-deposited $NbSe_3$ samples. The XRD patterns indicate the formation of a single-phase $NbSe_3$ wherein the lattice constants can be indexed to the monoclinic structure reported by the Joint Committee on Powder Diffraction Standards (JCPDS) card No. 75-0763.

Figure 3 shows high-resolution SEM, HRTEM, SAED, and TEM images of $NbSe_3$ nanoribbons prepared by our CVT method. The cross-sectional SEM image in figure 3(a) illustrates the nanoribbon-like morphology of the sample indicating a thickness of ~ 95 nm. The single-crystalline nature of the faceted nanoribbon structures is evidenced by the SAED pattern and lattice-resolved HRTEM image shown in figure 3(b)-(c). The SAED patterns do not change along the length of the nanoribbon, underscoring their single-crystalline nature. The fringe spacings observed in figure 3(c) are 0.313 and 0.471 nm, which can be indexed to the interplanar spacing between (012) and (200) planes, respectively. TEM images of the nanoribbons shown in figure 3(d, e, f) also enable the construction of size distribution histograms based on the dimensions of >100 individual nanoribbons (inset 3(e)). These analyses suggest a width distribution centered about 198 ± 53 nm in width, although nanoribbons with widths < 35 nm were also synthesized. The inset to Fig. 3e and Fig. 3f highlight the nanoribbon morphology, which makes the prepared nanostructures amenable to flexion under strain up to high angles without debonding and fracture.

We now turn to the electrical transport properties of our $NbSe_3$ nanoribbons. Device structures have been separately constructed for more than 12 individual nanoribbons. In $NbSe_3$, incommensurate charge-density-waves appear in two of the three types of trigonal pairings of its unit cell [18]: one along wave vector $q_1$ (0, 0.243 $\pm 0.005$, 0) directed across the main chain axis and the second in wave vector $q_2$ (0.5, 0.263 $\pm 0.005$, 0.5) [19]. This appearance is associated with a partial gap in the Fermi surface; in transport measurements, this Peierls transition is observed as a sharp increase in resistance at two different temperatures $T_{P1}$ and $T_{P2}$ [20]. We see this expected behaviour in our devices (a typical device is shown in figure 4(a) inset) by measuring resistance as a function of temperature in the four-terminal configuration. Two Peierls transitions at $T_{P1}$ and $T_{P2}$ are observed at 141 K and 59 K respectively. It is known that the residual resistance, $r_R$ (=R(300 K)/R(6 K)), decreases exponentially with decreasing thickness in the nanoscale regime [21] in $NbSe_3$. For our devices $r_R$ ranges from around 1.5 to 8 which is consistent with nanoscale devices seen elsewhere [10,15]. For bulk $NbSe_3$, $r_R$ = 220 is a typical value previously observed [21].

At sufficiently low temperatures the CDW will align itself to the distorted lattice by pinning to impurities. When an electric field is applied above a certain threshold, the charge carriers are depinned, and exhibit highly conductive, non-Ohmic behavior [22]. In order to observe such behaviour in our samples, we sum together an AC and DC signal to measure the differential resistance, dV/dI, as a function of electric field, E, in two-terminal configuration. Data shown in figure 4(b)-(c) were measured around the non-metallic regions after the second and first Peierls transitions respectively. Both plots show similar characteristics: Traces at or very near the Peierls transitions show little if any discernable evidence of pinning or depinning phenomenon. At temperatures just a few degrees cooler, the traces exhibit a plateau around E = 0 before showing a sharp drop in dV/dI above a critical value of electric field [23]. This resulting plateau represents the pinning of charge carriers to the impurity sites in the sample. The value of the electric field where the differential resistance drops represents the depinning of charge carriers, and is known as the threshold electric field, $E_T$. As we measure at lower temperatures, $E_T$ increases, but the corresponding drop in dV/dI decreases. Eventually the plateau feature is no longer observed at temperatures far from the Peierls transitions. The presence of large fluctuations in dV/dI just before depinning, particularly seen in the trace at 30 K, have been observed before in bulk samples [24-26]. It has been suggested that the origin of such noise may come from either configurational rearrangements of the CDW state or in general, CDW sliding motion closer to the threshold voltage [23, 26]. It has also been shown that the broad-band noise power spectrum in $NbSe_3$ is sample dimension dependent [28], however size effects (especially in the nanoscale regime) on such electric-field-driven noise phenomena have not been comprehensively explored, and is thus the focus for some of our future work.

Insets of figure 4(b)-(c) show plots of $E_T$ as a function of temperature around the second and around the first Peierls transitions respectively. We find that in our device the trend follows $E_T \sim e^{-T/T_o}$, where $T/T_o$ is associated with the average thermal fluctuations of the CDW order parameter [29]. This dependence has been previously observed not only in $NbSe_3$ but also in other correlated electron materials [30, 31]. We calculate $T_o$ to be 40 K around the first Peierls transition, and around 13 K for the second Peierls transition. These values are consistent with those found in bulk and in samples with nanoscale thicknesses [15, 21, 32] indicating that $T_o$ may be size independent. $E_T$ values range from 51 V/cm to 8.3 V/cm around the second Peierls transition, and from 46 V/cm to 23 V/cm around the first Peierls transition in our devices. These values are much higher than that found in bulk [30]; such magnitudes could be due to elevated number of defects or to the role of lattice reconstruction, especially at surfaces, upon scaling to finite size. In fact, previous studies on quasi two-dimensional $NbSe_3$ samples with nanoscale thicknesses have shown exponential increases in $E_T$ values with decreasing thicknesses [21, 33]. Moreover, it has recently been shown that $E_T$ values on the order of $10^1$ V/cm have been observed in quasi-one dimensional samples in which the dimensions are reduced along two directions, namely along the thickness and along the width [9, 34].

It is interesting to note, as also reported by Slot et al. [9], that $E_T$ at 120 K as a function of room temperature resistance per unit length, R/L, obeys a power law, $E_T \sim (R/L)^{2/3}$ when one-dimensional confinement is present in $NbSe_3$ samples. For our samples we report $E_T$ values at 120 K in the range of 13-26 V/cm, with corresponding R/L values of 391-173 Ω/μm. Comparing these values to the data in Fig. 1 of Ref. 9, our data is highly consistent, and fall within the trend line underscoring quasi-one-dimensional confinement in our samples.

**4. Conclusions**

Since charge-density-waves are present in quasi-one dimensional materials, fundamental studies of these properties are best undertaken on high-quality samples with reduced dimensions. Here, we have demonstrated a new useful CVT, which yields high-quality single-crystalline $NbSe_3$ nanoribbons in a single step starting from elemental Nb and Se precursors. High resolution SEM and TEM images underscore the ribbon-like shape of $NbSe_3$ with average cross sectional area ~ $10^4$ nm$^2$ and length ~ $10^1$ μm. The nanoribbons exhibit close to ideal stoichiometry, have uniformly rectangular cross-section and the prepared morphologies allow them to withstand high-angle deformations without mechanical fracture. From electrical transport measurements, our values of $r_R$ and of $E_T$ indicate nanoscale confinement; moreover, by comparing our values of $E_T$ with those from nanodevices measured previously in the literature we evidence that our nanoribbons exhibit characteristic quasi-one-dimensional signatures. In studying nanoscale samples, a better understanding of the physical mechanisms that modulate charge-density-waves can be achieved, and may provide increased insight to the mechanisms behind several other correlated electron systems. This work was primarily supported by the National Science Foundation grants under DMR 0847169 and DMR 0847324.

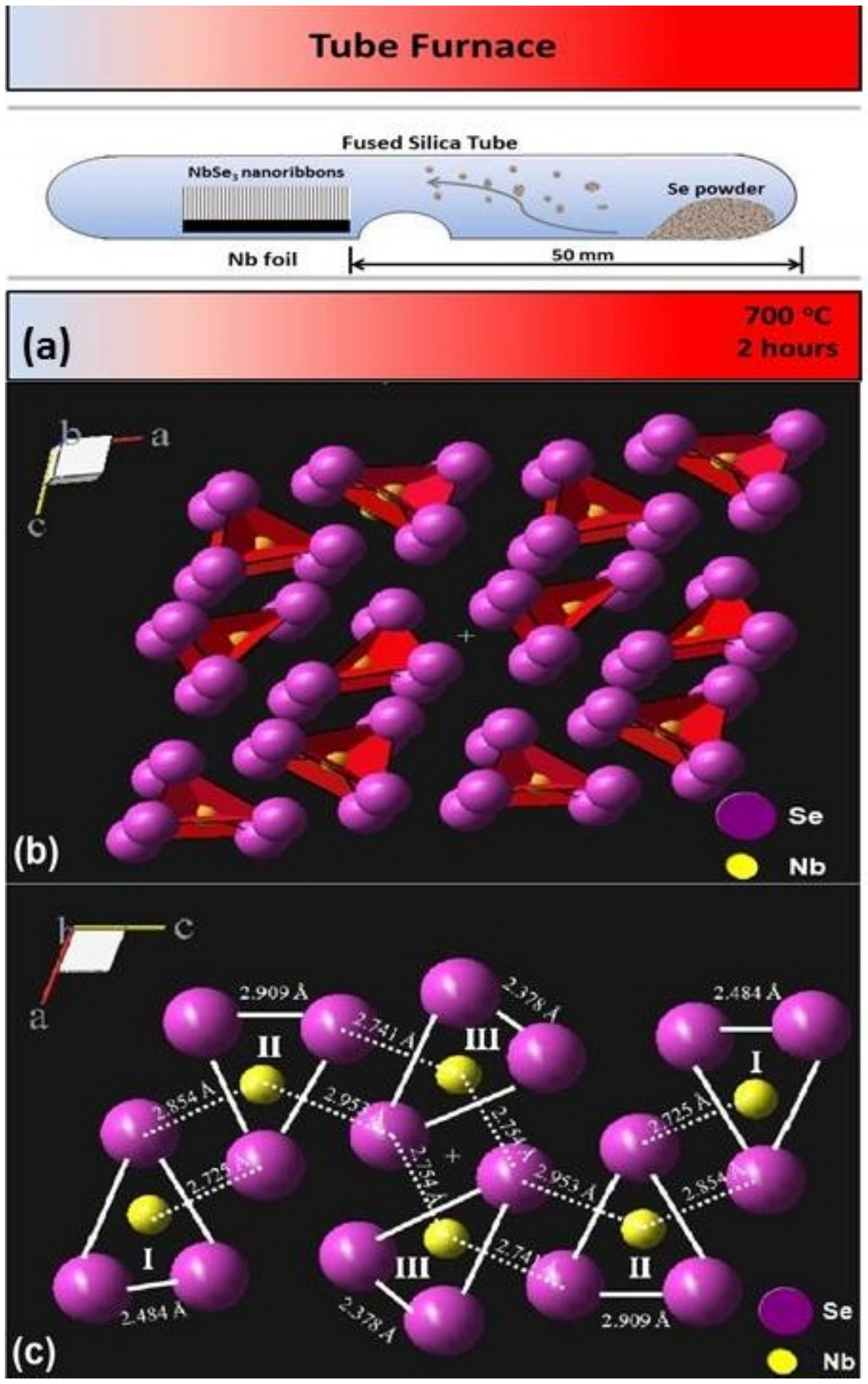

**Figure 1.** (a) Schematic depiction of the arrangement of the sealed quartz ampoule inside the walls of the horizontal tube furnace. (b) Illustration of the chainlike structure in $NbSe_3$. (c) Projection of the $NbSe_3$ unit cell perpendicular to the monoclinic *b* axis (out of the plane of the paper). The interchain Nb-Se and Se-Se bond lengths are depicted alongside the representation of the three symmetry-equivalent trigonal prism chain units.

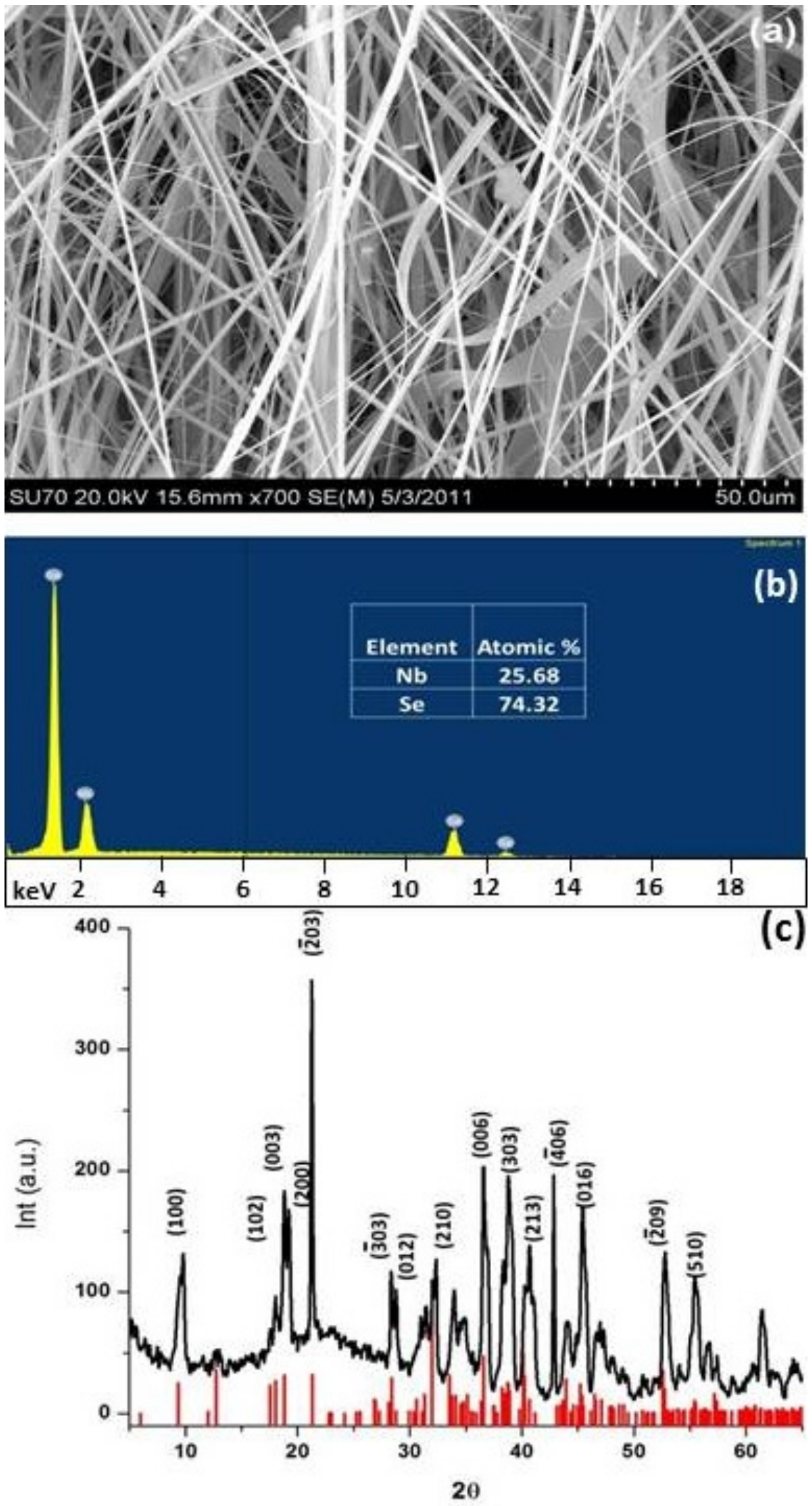

**Figure 2.** $NbSe_3$ nanoribbons synthesized by the CVT of Se powder onto a Nb substrate at 700°C for 2 hours. (a) SEM image depicting the nanoribbon-like morphology; (b) EDX spectrum acquired for the as-deposited substrate corroborating the stoichiometry of Nb and Se atoms; (c) X-ray diffraction pattern. The vertical lines indicate the peak positions expected for monoclinic $NbSe_3$ from JCPDS Card no. 75-0763.

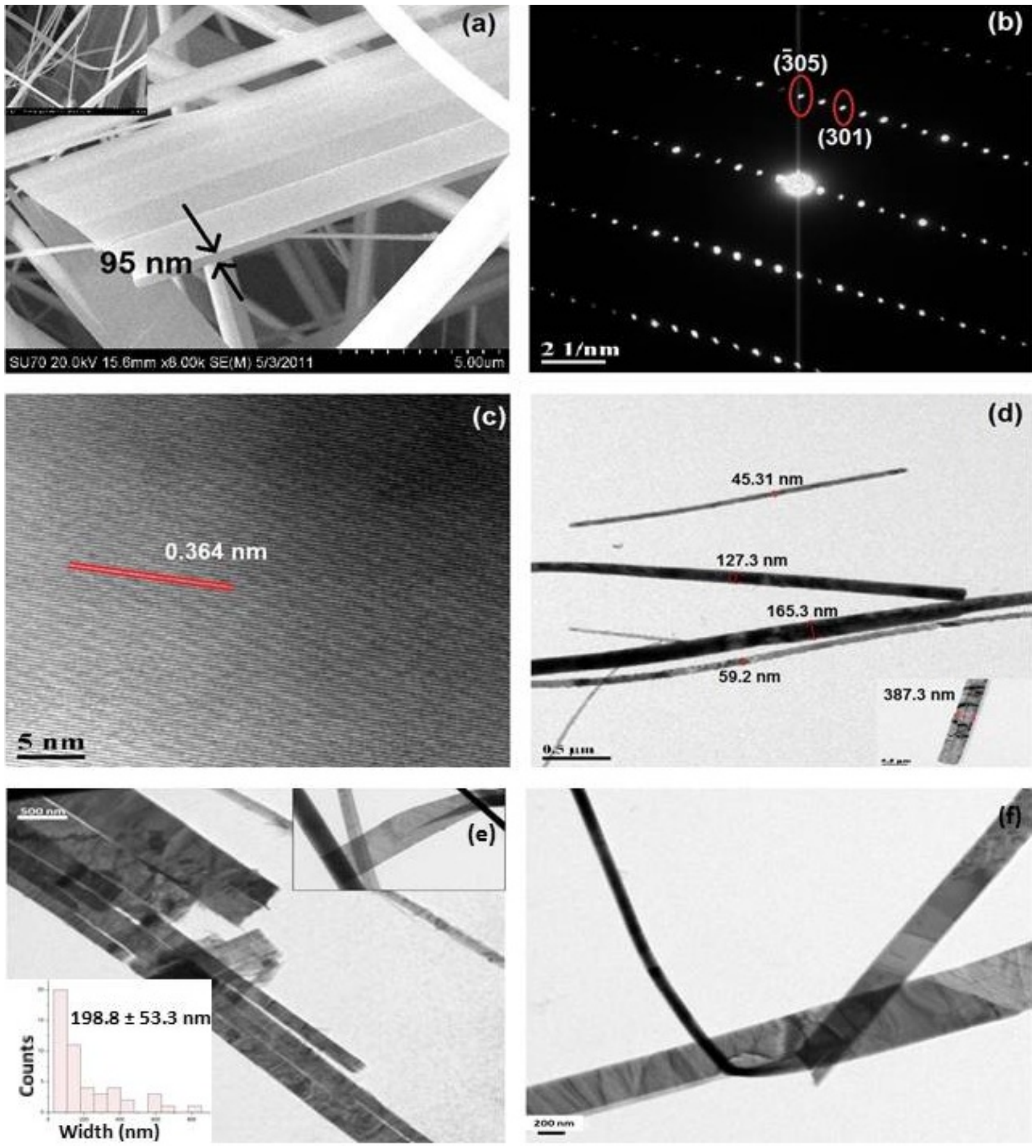

**Figure 3.** (a) Cross-sectional high-resolution scanning electron micrograph of the faceted ribbon-like nanostructure depicting the as-deposited $NbSe_3$ nanoribbons. Inset shows a SEM image of the sample. (b) Indexed SAED pattern for $NbSe_3$ nanoribbons (c) Lattice-resolved HRTEM image of an individual $NbSe_3$ nanoribbon (d, e, f) Low-magnification TEM images of several nanostructures highlighting the nanoribbon morphology and illustrating that the materials can be substantially bent up to high angles without fracture. Lower left inset showing the width distribution from the TEM images.

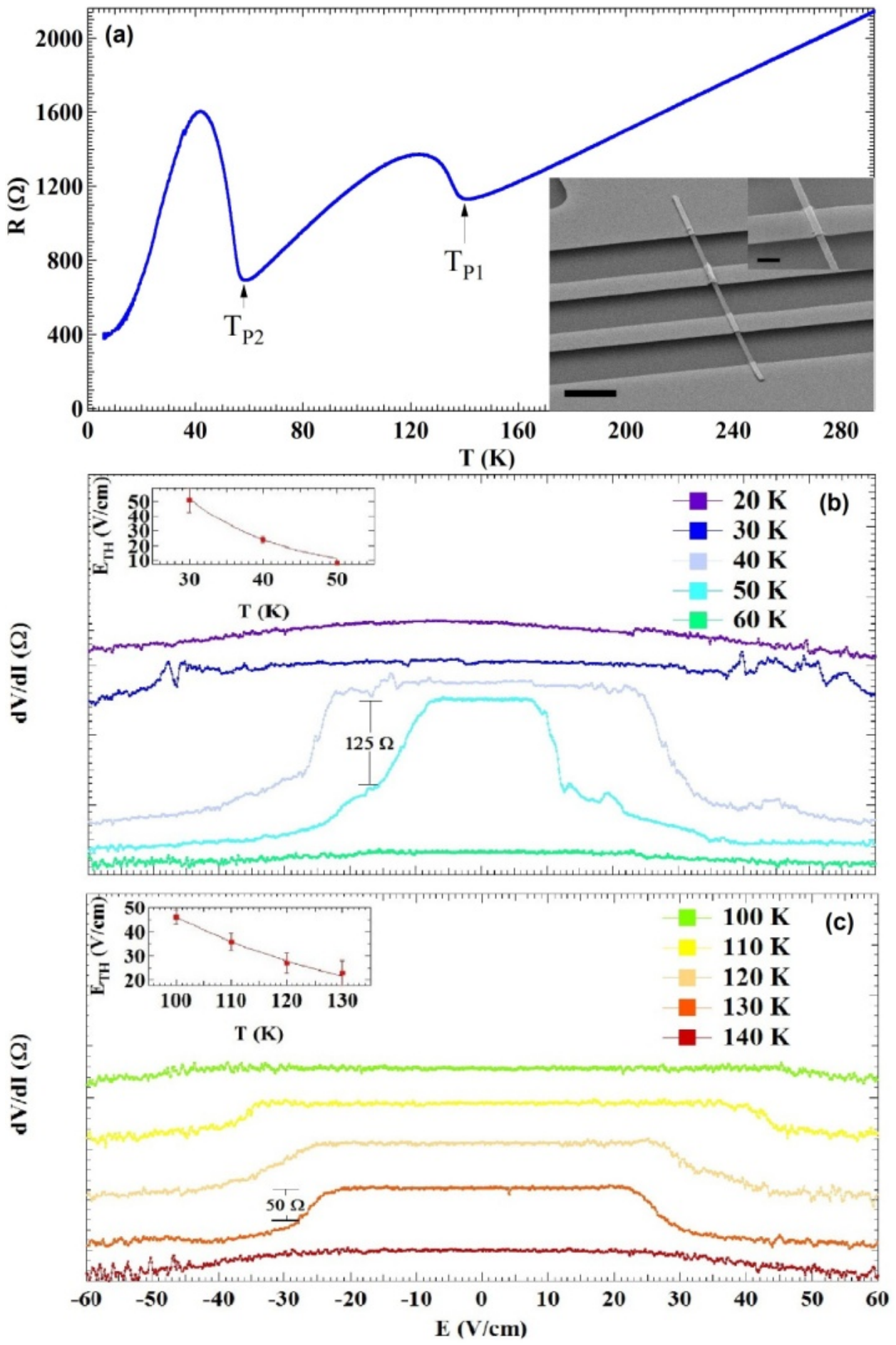

**Figure 4.** (a) Four-terminal resistance as a function of temperature of a single $NbSe_3$ nanoribbon. Two Peierls transitions occur at T = 141 K and 59 K. Excitation current = 100 nA. Inset: SEM images of a typical device. Scale bars for main image and for top left image are 5 µm and 1.5 µm respectively. (b) Two-terminal differential resistance vs electric field of a single $NbSe_3$ nanoribbon taken in the temperature region around $T_{P2}$. (c) Same measurement as done in (b) but taken in the temperature region around $T_{P1.}$ Traces in (b) and (c) are offset for clarity. Insets for (b) and (b) are plots of threshold electric field vs temperature. The fits are described in the text.